\documentclass[aps,pre,onecolumn,floatfix]{revtex4}

\usepackage{graphicx}
\usepackage{amssymb,amsfonts,amsmath}
\usepackage{epsf}
\usepackage{subfigure}
\usepackage{epstopdf}
\usepackage{mathrsfs}
\usepackage{color}

\newcommand{\be}{\begin{equation}}
\newcommand{\ee}{\end{equation}}
\newcommand{\bea}{\begin{eqnarray}}
\newcommand{\eea}{\end{eqnarray}}

\begin{document}

\title{Finite-time fluctuation theorem for diffusion-influenced surface reactions}

\author{Pierre Gaspard}
\email{gaspard@ulb.ac.be}
\affiliation{ Center for Nonlinear Phenomena and Complex Systems, Universit{\'e} Libre de Bruxelles (U.L.B.), Code Postal 231, Campus Plaine, B-1050 Brussels, Belgium}

\author{Raymond Kapral}
\email{rkapral@chem.utoronto.ca}
\affiliation{ Chemical Physics Theory Group, Department of Chemistry, University of Toronto, Toronto, Ontario M5S 3H6, Canada}

\begin{abstract}
A finite-time fluctuation theorem is proved for the diffusion-influenced surface reaction ${\rm A}\rightleftharpoons{\rm B}$ in a domain with any geometry where the species A and B undergo diffusive transport between the reservoir and the catalytic surface.  A corresponding finite-time thermodynamic force or affinity is associated with the symmetry of the fluctuation theorem.  The time dependence of the affinity and the reaction rates characterizing the stochastic process can be expressed analytically in terms of the solution of deterministic diffusion equations with specific boundary conditions.
\end{abstract}

\maketitle

\section{Introduction}

When driven out of equilibrium by thermodynamic forces or affinities, systems composed of atoms and molecules manifest macroscopic fluxes dissipating energy and producing thermodynamic entropy \cite{D36,P67,N79,GM84,C85}. In particular, for diffusion-influenced surface reactions, reactant and product molecules diffuse between the reservoir where they enter or exit the system and the catalytic surface where they undergo interconversion \cite{K81}.  On macroscopic scales, such processes are described by deterministic diffusion equations with suitable boundary conditions on the mean concentration fields of the reacting species.  However, on mesoscopic scales, molecular motion is erratic and reactive events occur at random on the catalytic surface, which requires a description in terms of stochastic processes \cite{N72,G04}.  As a matter of principle, these processes should be compatible with the underlying microscopic Hamiltonian mechanics and its microreversibility.  Consequently, time-reversal symmetry relations, known as fluctuation theorems \cite{ECM93,G96,K98,LS99,J11,G13NJP}, are satisfied by the fluctuations of the currents flowing across nonequilibrium systems.  These theorems are formulated within the framework of large-deviation theory \cite{T09} since they concern the full counting statistics of the currents, including rare events that are exponentially suppressed in time.  For systems in stationary states, fluctuation theorems are time-reversal symmetry relations holding in the long-time limit.  This has been established, in particular, for systems sustaining reactions or transport by diffusion \cite{G04JCP,AG04,AG06,DDR04,D07,SIPGD07,PGPD10,PGLD13,BDGJL15}.

However, it has been shown in Ref.~\cite{AG08} that such time-reversal symmetry relations may also hold over finite time intervals for certain reactions taking place in systems without spatial extension.  In these systems, a thermodynamic force or affinity can thus be defined at every time as a consequence of the finite-time symmetry.

Here, our purpose is to show that such a finite-time fluctuation theorem also holds in spatially extended systems where a surface reaction is controlled by the diffusion of reactants and products from and to the reservoir.  The problem is formulated within the theory of stochastic partial differential equations in terms of stochastic diffusion equations coupled by stochastic boundary conditions for the reaction ${\rm A}\rightleftharpoons{\rm B}$ at the catalytic surface.  In this framework, a finite-time fluctuation theorem is established for the probability distribution that a certain number of reactive events have occurred during some finite time interval. The theorem is proved by spatial discretization into small cells, leading to a Markov jump process ruling the time evolution of the numbers of molecules inside the cells.  The master equation of this Markov jump process can be exactly solved using the generating function method~\cite{G04}, which provides the analytical expression for the cumulant generating function at every time.  Returning to the continuum description, the cumulant generating function is obtained in terms of finite-time rates given by solving deterministic diffusion equations with specific boundary conditions.  The large-deviation properties of the spatially extended stochastic process can thus be found by solving deterministic partial differential equations.

The paper is organized as follows.  The main result is presented in Sec.~\ref{Sec-thm} where the finite-time fluctuation theorem is stated for the probability distribution of the number of reactive events and the associated cumulant generating function.  In this section, the finite-time rates and the corresponding affinity are expressed in terms of the solution of deterministic diffusion equations with the specific boundary conditions, and connection is made to the thermodynamic entropy production.  The proof of the finite-time fluctuation theorem is carried out in Sec.~\ref{Sec-proof}.  Section~\ref{conclusion} gives concluding remarks and perspectives.

\section{The main results}
\label{Sec-thm}

\subsection{Stochastic partial differential equations for the diffusion-influenced surface reaction}

Let us consider a diffusive medium of dimension $d$ and volume $V$, extending between three surfaces $\partial V = S_{\rm cat}\cup S_{\rm inert}\cup S_{\rm res}$.  $S_{\rm cat}$~is a catalytic surface where the reaction ${\rm A}\rightleftharpoons{\rm B}$ takes place.  $S_{\rm inert}$ is an inert surface where the species ${\rm A}$ and ${\rm B}$ are reflected.  $S_{\rm res}$ is a surface in contact with a reservoir for the species ${\rm A}$ and ${\rm B}$. These species undergo diffusion in the volume $V$ so that their concentrations, $c_{\rm A}$ and~$c_{\rm B}$, obey the stochastic diffusion equations,
\bea
&& \partial_t c_{\rm A} + \pmb{\nabla}\cdot {\bf j}_{\rm A} = 0 \, ,\qquad {\bf j}_{\rm A}  = -D_{\rm A} \pmb{\nabla} c_{\rm A} + \pmb{\eta}_{\rm A}\, , \label{diff-eq-A}\\
&& \partial_t c_{\rm B} + \pmb{\nabla}\cdot {\bf j}_{\rm B} = 0 \, ,\qquad {\bf j}_{\rm B}  = -D_{\rm B}\pmb{\nabla} c_{\rm B} + \pmb{\eta}_{\rm B} \, , \label{diff-eq-B}
\eea
expressed in terms of Gaussian noise fields such that
\be
\langle\pmb{\eta}_k({\bf r},t)\rangle = 0 \, , \qquad\qquad \langle\pmb{\eta}_k({\bf r},t)\otimes\pmb{\eta}_{k'}({\bf r'},t')\rangle = 2\,  D_k \, c_k({\bf r}, t) \, \delta_{kk'} \, \delta({\bf r}-{\bf r'}) \, \delta(t-t') \, {\boldsymbol{\mathsf 1}}
\label{Gaussian-eta}
\ee
for $k,k'={\rm A},{\rm B}$, where $D_k$ are positive diffusion coefficients and ${\boldsymbol{\mathsf 1}}$ is the $d\times d$ identity matrix. The boundary conditions are given by
\bea
{\rm if}\ {\bf r}\in S_{\rm cat} : \qquad  && D_{\rm A} \, \partial_{\bot} c_{\rm A}({\bf r},t) = -D_{\rm B}\, \partial_{\bot}c_{\rm B}({\bf r},t) =\kappa_+ c_{\rm A}({\bf r},t) - \kappa_- c_{\rm B}({\bf r},t)+ \xi({\bf r},t) \, , \label{bc-cat}\\
{\rm if}\ {\bf r}\in S_{\rm inert} : \qquad && \quad\ \ \partial_{\bot} c_{\rm A}({\bf r},t) = 0 \, , \quad \ \partial_{\bot}c_{\rm B}({\bf r},t) = 0\, , \label{bc-inert}\\
{\rm if}\ {\bf r}\in S_{\rm res} : \qquad && \quad\ \ c_{\rm A}({\bf r},t) = \bar{c}_{\rm A} \, , \qquad c_{\rm B}({\bf r},t) = \bar{c}_{\rm B}\, , \label{bc-res}
\eea
where $\partial_{\bot}={\bf 1}_{\bot}\cdot\pmb{\nabla}$ is the gradient in the direction of the unit vector ${\bf 1}_{\bot}$ normal to the surface and oriented towards the interior of the volume $V$, and $\kappa_{\pm}$ are the positive rate constants of the surface reactions. These rate constants have the SI units of meter per second.  $\bar{c}_{\rm A}$ and~$\bar{c}_{\rm B}$ denote the given concentrations at the reservoir.  The Gaussian noise field due to the surface reaction is characterized by
\be
\langle \xi({\bf r},t) \rangle = 0 \, , \qquad\qquad
\delta^{\rm s}({\bf r})\, \langle \xi({\bf r},t)\,\xi({\bf r'},t') \rangle\, \delta^{\rm s}({\bf r'}) = (\kappa_+\, c_{\rm A}+\kappa_-\, c_{\rm B}) \, \delta^{\rm s}({\bf r}) \, \delta({\bf r}-{\bf r'}) \, \delta(t-t')  \, ,
\label{Gaussian-xi}
\ee
in terms of surface delta distributions $\delta^{\rm s}({\bf r})$, nonvanishing if ${\bf r}\in S_{\rm cat}$ \cite{BAM76}.

\subsection{The finite-time fluctuation theorem}
\label{FTFT}

Let the random variable $n$ denotes the number of reactive events ${\rm A}\to{\rm B}$ that have occurred during the time interval $[0,t]$, if the system is in a steady state with given concentrations $\bar{c}_{\rm A}$ and $\bar{c}_{\rm B}$ at the reservoir.  The probability $P(n,t)$ that $n$ reactive events have occurred is equal to
\be
P(n,t) = {\rm e}^{-t\left(W_t^{(+)}+W_t^{(-)}\right)} \left(\frac{W_t^{(+)}}{W_t^{(-)}}\right)^{n/2} I_n\left(2t\sqrt{W_t^{(+)}W_t^{(-)}}\right) ,
\label{P(n)}
\ee
where $W_t^{(\pm)}$ are two finite-time rates explicitly given below and $I_n(u)$ is the modified regular Bessel function defined in Sec.~9.6 of Ref.~\cite{AS72}.  Since $I_n(u)=I_{-n}(u)$, this probability distribution obeys the finite-time fluctuation theorem
\be
\boxed{\frac{P(n,t)}{P(-n,t)} = \exp ( {\cal A}_t \, n )}
\label{FT}
\ee
holding {\it at every time} with the finite-time affinity defined as
\be
{\cal A}_t = \ln\frac{W_t^{(+)}}{W_t^{(-)}} .
\label{A-dfn}
\ee
The finite-time rates have the explicit forms
\bea
&&W_t^{(+)} = \Sigma \kappa_+ \bar{c}_{\rm A} + \frac{1}{t} \, \Psi(t) \, , \label{W+} \\
&&W_t^{(-)} = \Sigma \kappa_- \bar{c}_{\rm B} + \frac{1}{t} \, \Psi(t) \,  , \label{W-}
\eea
with
\be
\Psi(t) = \ell^2 \kappa_+\kappa_- \left[ \frac{\bar{c}_{\rm B}}{D_{\rm A}^2} \Upsilon_{\rm A}(t) + \frac{\bar{c}_{\rm A}}{D_{\rm B}^2} \Upsilon_{\rm B}(t)\right] .
\label{Psi}
\ee
The first terms on the right sides of Eqs.~(\ref{W+}) and~(\ref{W-}) are proportional to the effective catalytic surface area
\be
\Sigma = \int_{\rm cat} dS \, (1-\phi) \, ,
\label{Sigma}
\ee
$\phi$ being the solution of the following stationary problem,
\bea
&&\nabla^2 \phi = 0 \, , \label{phi1}\\
&& \left(\partial_{\bot} \phi\right)_{\rm cat} = \ell^{-1}(\phi-1)_{\rm cat} \, , \label{phi2}\\
&& \left(\partial_{\bot} \phi\right)_{\rm inert} = 0 \, , \label{phi3}\\
&& \left(\phi\right)_{\rm res} = 0 \, , \label{phi4}
\eea
where
\be
\ell \equiv \left(\frac{\kappa_+}{D_{\rm A}} + \frac{\kappa_-}{D_{\rm B}}\right)^{-1}
\label{ell}
\ee
is the characteristic length of the diffusion-influenced surface reaction.
In Eq.~(\ref{Psi}), $\Psi(t)$ is given in terms of the time-dependent functions
\be
\Upsilon_k(t) = \int dV \phi({\bf r})\left[ \phi({\bf r})-f_k({\bf r},t)\right] \, ,
\label{Upsilon}
\ee
where $f_k$ is the solution of the following time-dependent problem,
\bea
&&\partial_t f_k = D_k \nabla^2 f_k \, , \label{f1}\\
&& \left(\partial_{\bot} f_k\right)_{\rm cat} = \left(\frac{\kappa_+}{D_{\rm A}}\, f_{\rm A}+\frac{\kappa_-}{D_{\rm B}}\, f_{\rm B}\right)_{\rm cat} \, , \label{f2}\\
&& \left(\partial_{\bot} f_k\right)_{\rm inert} = 0 \, , \label{f3}\\
&& \left(f_k\right)_{\rm res} = 0 \, , \label{f4}\\
&& \left(f_k\right)_{t=0} = \phi \, , \label{f5}
\eea
for $k={\rm A},{\rm B}$.

If the catalytic and inert surfaces $S_{\rm inert}\cup S_{\rm cat}$, as well as the domain $V$, are compact the constant $\Sigma$ and the functions $\Upsilon_k(t)$ are bounded, so that the rates~(\ref{W+}) and~(\ref{W-}) converge in the long-time limit $t\to\infty$ to their asymptotic values
\bea
&&W_{\infty}^{(+)} = \Sigma \kappa_+ \bar{c}_{\rm A}  \, , \label{Winf+} \\
&&W_{\infty}^{(-)} = \Sigma \kappa_- \bar{c}_{\rm B}  \,  , \label{Winf-}
\eea
whereupon the affinity~(\ref{A-dfn}) converges to the finite value
\be
{\cal A}_{\infty} = \ln\frac{W_{\infty}^{(+)}}{W_{\infty}^{(-)}}= \ln\frac{\kappa_+\bar{c}_{\rm A}}{\kappa_-\bar{c}_{\rm B}} .
\label{A-inf}
\ee
We expect the same behavior to hold if the catalytic and inert surfaces $S_{\rm inert}\cup S_{\rm cat}$ are compact and delimit a finite volume, while the domain $V$ is non-compact, but three-dimensional.

\subsection{Formulation in terms of the cumulant generating function}
\label{CGF}

Introducing the cumulant generating function
\be
Q_t(\lambda) \equiv  -\frac{1}{t} \ln \sum_{n=-\infty}^{+\infty} {\rm e}^{-\lambda n} P(n,t)
\label{Q-dfn}
\ee
with the counting parameter $\lambda$, we have the result that
\be
Q_t(\lambda) = W_t^{(+)}\left(1-{\rm e}^{-\lambda}\right) +  W_t^{(-)}\left(1-{\rm e}^{\lambda}\right) ,
\label{Q}
\ee
where the finite-time rates $W_t^{(\pm)}$ were defined in Eqs.~(\ref{W+}) and~(\ref{W-}).  As a consequence of the finite-time fluctuation theorem~(\ref{FT}), the following symmetry relation is satisfied {\it at every time},
\be
Q_t(\lambda) = Q_t({\cal A}_t-\lambda),
\label{Q-sym}
\ee
in terms of the finite-time affinity~(\ref{A-dfn}).  The mean current and the diffusivity at time $t$ are thus given by
\bea
&& {\cal J}_t =\frac{\partial Q_t}{\partial\lambda}(0) = W_t^{(+)} - W_t^{(-)} \, , \label{J_t}\\
&& {\cal D}_t =-\frac{1}{2}\frac{\partial^2 Q_t}{\partial\lambda^2}(0) = \frac{1}{2} \left(W_t^{(+)} +W_t^{(-)}\right) \, . \label{D_t}
\eea
We notice that the mean current~(\ref{J_t}) does not depend on time because of the forms of the expressions~(\ref{W+}) and~(\ref{W-}) for the rates, implying
\be
{\cal J}_t = {\cal J} =W_{\infty}^{(+)}-W_{\infty}^{(-)}= \Sigma\left( \kappa_{+}\bar{c}_{\rm A} - \kappa_{-}\bar{c}_{\rm B} \right) .
\label{J}
\ee

Moreover, the stationary solutions for the species concentrations are given by
\be
\langle c_k\rangle_{\rm st} = \bar{c}_k +\frac{\nu_k\ell}{D_k}   \left(\kappa_+\bar{c}_{\rm A}-\kappa_- \bar{c}_{\rm B}\right)  \phi \qquad (k={\rm A},{\rm B}),
\label{NESS}
\ee
where $\phi$ is the solution of the problem~(\ref{phi1})-(\ref{phi4}), while $\nu_{\rm A}=-1$ and $\nu_{\rm B}=+1$ are the stoichiometric coefficients of the reaction ${\rm A}\to{\rm B}$.  The equilibrium thermodynamic state occurs when the chemical equilibrium condition $\kappa_+\bar{c}_{\rm A}=\kappa_- \bar{c}_{\rm B}$ is satisfied, in which case the state is uniform.  The stationary solution~(\ref{NESS}) determines the cumulant generating function at early time according to
\be
Q_t(\lambda) = \int_{\rm cat} dS \left[ \kappa_+\langle c_{\rm A}\rangle_{\rm st} \left( 1-{\rm e}^{-\lambda}\right) + \kappa_- \langle c_{\rm B}\rangle_{\rm st} \left( 1-{\rm e}^{\lambda}\right) \right] +O(t),
\label{Q-early}
\ee
up to corrections that are linear in time.

The finite-time fluctuation theorem~(\ref{FT}) and the associated results, which are stated above, are proved in Sec.~\ref{Sec-proof} by extending the result obtained in Ref.~\cite{AG08}.

\subsection{Thermodynamic entropy production}
\label{Thermo}

Here, we show the equivalence between the expressions for the entropy production given by macroscopic nonequilibrium thermodynamics and the fluctuation theorem under stationary conditions.  On the one hand, according to nonequilibrium thermodynamics \cite{P67,GM84,N79}, the entropy production is equal to the sum of the contributions from diffusion in the domain $V$ and reaction at the catalytic surface $S_{\rm cat}$,
\be
\frac{1}{k_{\rm B}}\frac{d_{\rm i}S}{dt} = \int_V dV \left[ D_{\rm A} \frac{(\pmb{\nabla}a)^2}{a} + D_{\rm B} \frac{(\pmb{\nabla}b)^2}{b}\right] +\int_{\rm cat} dS \left(\kappa_+ a-\kappa_-b\right) \ln\frac{\kappa_+ a}{\kappa_-b} \geq 0
\ee
with the notations $a=\langle c_{\rm A}\rangle_{\rm st}$ and $b=\langle c_{\rm B}\rangle_{\rm st}$.  Now, we have that
\be
\int_V dV \frac{(\pmb{\nabla}a)^2}{a} = \int_V dV \, \pmb{\nabla}a\cdot\pmb{\nabla}\left(\ln\kappa_+ a\right)
= \int_{\partial V} d{\bf S} \cdot(\pmb{\nabla}a) \, \ln\kappa_+ a\, ,
\ee
by using the divergence theorem and the fact that $\nabla^2a=0$ in a steady state.  A similar expression is obtained for the other concentration field $b$.  The boundary $\partial V$ of the domain $V$ is composed of the catalytic, inert, and reservoir surface components, where $d{\bf S}=-{\bf 1}_{\bot} dS$ if ${\bf 1}_{\bot}$ is the unit vector normal to the surface and oriented towards the interior of the domain.  Using the boundary conditions~(\ref{bc-cat})-(\ref{bc-res}), we find that the contributions of the catalytic and inert components of the surface cancel and there remain the contributions of the surface component in contact with the reservoir,
\be
\frac{1}{k_{\rm B}}\frac{d_{\rm i}S}{dt} = \int_{\rm res} dS\, (-D_{\rm A}\, \partial_{\bot} a) \, \ln\kappa_+ a +\int_{\rm res} dS\, (-D_{\rm B}\, \partial_{\bot} b) \, \ln\kappa_- b \, .
\ee
Replacing the concentration fields at the reservoir by their expression~(\ref{NESS}) in terms of the field $\phi$ obeying Eqs.~(\ref{phi1})-(\ref{phi4}), and using the fact that
\be
0=\int_V dV \, \nabla^2\phi = \int_{\partial V} d{\bf S}\cdot\pmb{\nabla}\phi = \frac{1}{\ell} \, \Sigma -\int_{\rm res} dS \, \partial_{\bot}\phi \, ,
\ee
where $\Sigma$ is defined in~Eq.~(\ref{Sigma}), we find that the entropy production is given by
\be
\frac{1}{k_{\rm B}}\frac{d_{\rm i}S}{dt} = \Sigma \left(\kappa_+ \bar{c}_{\rm A}-\kappa_- \bar{c}_{\rm B}\right) \ln\frac{\kappa_+\bar{c}_{\rm A}}{\kappa_-\bar{c}_{\rm B}} ={\cal J} {\cal A}_{\infty} \geq 0 \, ,
\label{entrprod-macro}
\ee
which is determined by the reservoir values of the concentrations.  Therefore, the entropy production is equal to the mean current~(\ref{J}) multiplied by the asymptotic value~(\ref{A-inf}) of the affinity, as expected.

On the other hand, the thermodynamic entropy production can be expressed as
\be
\frac{1}{k_{\rm B}}\frac{d_{\rm i}S}{dt} = \lim_{t\to\infty} \frac{1}{t} \sum_{n=-\infty}^{+\infty} P(n,t) \, \ln \frac{P(n,t)}{P(-n,t)} = \lim_{t\to\infty} {\cal J}  {\cal A}_t ={\cal J}  {\cal A}_{\infty} \geq 0
\label{entrprod-FT}
\ee
in terms of the probability distribution~(\ref{P(n)}).  Since this latter obeys the fluctuation theorem~(\ref{FT}), we recover the macroscopic value~(\ref{entrprod-macro}) of the entropy production because the mean current in the steady state is given by ${\cal J}=\langle n(t)\rangle_{\rm st}/t$ according to Eq.~(\ref{J}).  The fluctuation theorem is thus consistent with macroscopic nonequilibrium thermodynamics for the diffusion-influenced surface reaction.  In addition, the non-negative quantity ${\cal J} {\cal A}_t\geq 0$ in Eq.~(\ref{entrprod-FT}) can be interpreted as a finite-time entropy production in the measurement of the surface reaction using the full counting statistics of the reactive events.


\section{Proof of the finite-time fluctuation theorem}
\label{Sec-proof}

The proof of the finite-time fluctuation theorem~(\ref{FT}) and the associated results stated in Subsecs.~\ref{FTFT} and~\ref{CGF} is carried out by discretizing space into small cells and using the master equation of the stochastic process for the random numbers of molecules in the cells, which, in the continuum limit, is equivalent to the stochastic process ruled by Eqs.~(\ref{diff-eq-A})-(\ref{Gaussian-xi}).  We solve this master equation using a method based on the moment generating function for the probability distribution of the molecular numbers and the number of reactive events occurring during some time interval \cite{G04,AG08}.  Since the kinetic equations for the mean values of the molecular numbers are linear, the steady state of the reaction-diffusion process is described by a Poisson distribution and the partial differential equation ruling the moment generating function admits an exact solution, yielding an expression for the cumulant generating function of the number of reactive events occurring during some time interval.  Its dependence on the counting parameter is obtained by using projectors onto the subspaces corresponding to the molecules of species A and B, and we find that the cumulant generating function has the form~(\ref{Q}).  As a consequence, the probability distribution of the random number of reactive events is given by Eq.~(\ref{P(n)}), implying the finite-time fluctuation theorem~(\ref{FT}).  In this discrete-space formulation, matricial expressions are obtained for the time-dependent rates.  

Returning to the continuum limit, we first show that we recover the macroscopic diffusion-reaction equations for the mean concentration fields.  Next, we deduce the analytical expressions for the time-dependent rates by transforming the matricial equations obtained by space discretization into partial differential equations and their boundary conditions.  This is performed by summing the matricial equations with arbitrary conjugate vectors in order to obtain expressions involving integrals in the continuum limit.  This method allows us to obtain the partial differential equations and their boundary conditions by considering variations with respect to the space-dependent conjugate functions corresponding in the continuum limit to the aforementioned arbitrary conjugate vectors.  In this way, the time-dependent rates are shown to be given by the solutions of the problems (\ref{phi1})-(\ref{phi4}) and (\ref{f1})-(\ref{f5}), finally yielding their analytical expressions~(\ref{W+})-(\ref{W-}) with the time-dependent function~(\ref{Psi}) and the constant~(\ref{Sigma}).  Also, expression~(\ref{Q-early}) is obtained for the behavior of the cumulant generating function at early time.

\subsection{Space discretization}

\subsubsection{Master equation}

In order to prove the theorem, the $d$-dimensional volume $V$ is discretized into small cubic cells $\{{\cal C}_{\bf r}\}$ of side $\Delta r$, volume $\Delta r^d$, and centered on the nodes ${\bf r}$ of a $d$-dimensional cubic lattice.  Every cell contains a certain number of molecules of each species:
\be
A_{\bf r} = \int_{{\cal C}_{\bf r}} c_{\rm A}({\bf r'}) \, d{\bf r'} \qquad\mbox{and} \qquad B_{\bf r} = \int_{{\cal C}_{\bf r}} c_{\rm B}({\bf r'}) \, d{\bf r'} \, .
\ee
Some of the cells are in contact with the catalytic surface, the inert surface, and the reservoir.  Every cubic cell has $2d$ faces, which correspond to the $2d$ vectors
\be
\Delta{\bf r} \in \left\{ (\pm\Delta r,0,\dots,0), (0,\pm\Delta r,\dots,0), \dots, (0,0,\dots,\pm\Delta r) \right\} ,
\ee
joining the center of the cell to those of the next-neighboring cells.  The cells in the bulk of the volume have all their faces in contact with next-neighboring cells.  However, the other cells have some faces in contact with the catalytic surface, the inert surface, or the reservoir.  Therefore, for every cell, the set of $2d$ vectors is subdivided as
\be
\left\{ \Delta{\bf r}\right\} = \left\{ \Delta{\bf r}\right\}_{\rm diff} \cup \left\{ \Delta{\bf r}\right\}_{\rm cat} \cup\left\{ \Delta{\bf r}\right\}_{\rm inert} \cup\left\{ \Delta{\bf r}\right\}_{\rm res}
\ee
into faces, through which particles can be exchanged by diffusion with next-neighboring cells or the reservoir, reflected on the inert surface, or transformed by reaction on the catalytic surface.

The molecular numbers change in time according to the following processes:
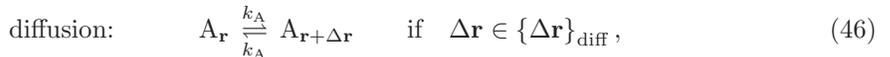
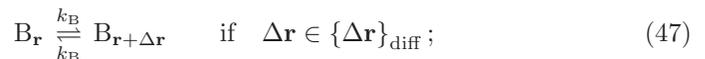
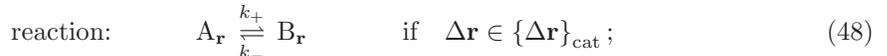
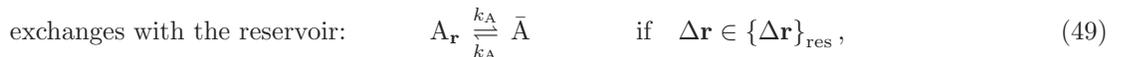
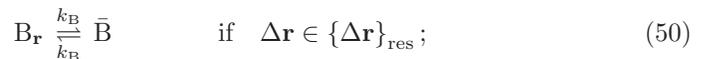
\bea
\mbox{diffusion:} \qquad &&{\rm A}_{\bf r} \; \mathop{\rightleftharpoons}^{k_{\rm A}}_{k_{\rm A}} \; {\rm A}_{{\bf r}+\Delta{\bf r}} \qquad \mbox{if}\quad \Delta{\bf r}\in \left\{ \Delta{\bf r}\right\}_{\rm diff}  ,\\
&&{\rm B}_{\bf r} \; \mathop{\rightleftharpoons}^{k_{\rm B}}_{k_{\rm B}} \; {\rm B}_{{\bf r}+\Delta{\bf r}} \qquad \mbox{if}\quad \Delta{\bf r}\in \left\{ \Delta{\bf r}\right\}_{\rm diff}  ;\\
\mbox{reaction:} \qquad &&{\rm A}_{\bf r} \; \mathop{\rightleftharpoons}^{k_+}_{k_-} \; {\rm B}_{\bf r} \qquad\quad\ \; \mbox{if}\quad \Delta{\bf r}\in \left\{ \Delta{\bf r}\right\}_{\rm cat}  ;\\
\mbox{exchanges with the reservoir:} \qquad &&{\rm A}_{\bf r} \; \mathop{\rightleftharpoons}^{k_{\rm A}}_{k_{\rm A}} \; \bar{\rm A} \qquad\qquad  \mbox{if}\quad \Delta{\bf r}\in \left\{ \Delta{\bf r}\right\}_{\rm res} , \\
&&{\rm B}_{\bf r} \; \mathop{\rightleftharpoons}^{k_{\rm B}}_{k_{\rm B}} \; \bar{\rm B} \qquad\qquad  \mbox{if}\quad \Delta{\bf r}\in \left\{ \Delta{\bf r}\right\}_{\rm res} ;
\eea
and there is no change at the faces in contact with the inert surface.  The rate constants are given by
\be
k_{\rm A} = \frac{D_{\rm A}}{\Delta r^2} \, , \qquad k_{\rm B} = \frac{D_{\rm B}}{\Delta r^2} \, , \qquad\mbox{and}\qquad k_{\pm} = \frac{\kappa_{\pm}}{\Delta r}
\label{rates}
\ee
in terms of the diffusion coefficients and surface rate constants of the continuous-space formulation.
These rate constants are positive and have the SI units of (second)$^{-1}$.

We consider the time evolution of the probability
\be
P=P\left(n,\{A_{\bf r}\}, \{B_{\bf r}\},t\right)
\ee
that the cells contain given molecular numbers and that $n$ reactive events have occurred during the time interval $[0,t]$.  This probability is ruled by the following master equation,
\bea
\frac{dP}{dt} =\hat{L}P &=& \sum_{\bf r}\biggl\{ \sum_{\left\{ \Delta{\bf r}\right\}_{\rm diff}} k_{\rm A} \left({\rm e}^{-\partial_{A_{{\bf r}+\Delta{\bf r}}}}{\rm e}^{+\partial_{A_{\bf r}}} - 1 \right) A_{\bf r} P \nonumber\\
&& \ \quad + \sum_{\left\{ \Delta{\bf r}\right\}_{\rm diff}} k_{\rm B} \left({\rm e}^{-\partial_{B_{{\bf r}+\Delta{\bf r}}}}{\rm e}^{+\partial_{B_{\bf r}}} - 1 \right) B_{\bf r} P \nonumber\\
&& \ \quad + \sum_{\left\{ \Delta{\bf r}\right\}_{\rm cat}} k_+ \left({\rm e}^{-\partial_n}{\rm e}^{+\partial_{A_{\bf r}}}{\rm e}^{-\partial_{B_{\bf r}}} - 1 \right) A_{\bf r} P \nonumber\\
&& \ \quad + \sum_{\left\{ \Delta{\bf r}\right\}_{\rm cat}} k_- \left({\rm e}^{+\partial_n}{\rm e}^{-\partial_{A_{\bf r}}}{\rm e}^{+\partial_{B_{\bf r}}} - 1 \right) B_{\bf r} P \nonumber\\
&& \ \quad + \sum_{\left\{ \Delta{\bf r}\right\}_{\rm res}} \left[ k_{\rm A}\bar{A} \left({\rm e}^{-\partial_{A_{\bf r}}} - 1 \right) P + k_{\rm A} \left({\rm e}^{+\partial_{A_{\bf r}}} - 1 \right) A_{\bf r} P\right] \nonumber\\
&& \ \quad + \sum_{\left\{ \Delta{\bf r}\right\}_{\rm res}} \left[ k_{\rm B}\bar{B} \left({\rm e}^{-\partial_{B_{\bf r}}} - 1 \right) P + k_{\rm B} \left({\rm e}^{+\partial_{B_{\bf r}}} - 1 \right) B_{\bf r} P\right]\biggr\} ,
\eea
where $\bar{A}=\bar{c}_{\rm A}\Delta r^d$ and $\bar{B}=\bar{c}_{\rm B}\Delta r^d$.

\subsubsection{Kinetic equations for the mean numbers}

As a consequence, the time evolution of the mean numbers,
\bea
\langle A_{\bf r}\rangle &=& \sum_{n,\{A_{\bf r}\}, \{B_{\bf r}\}} A_{\bf r} P \, , \\
\langle B_{\bf r}\rangle &=& \sum_{n,\{A_{\bf r}\}, \{B_{\bf r}\}} B_{\bf r} P \, , \\
\langle n \rangle &=& \sum_{n,\{A_{\bf r}\}, \{B_{\bf r}\}} n\, P \, ,
\eea
is ruled by the following equations,
\bea
\frac{d}{dt} \langle A_{\bf r}\rangle &=& \sum_{\left\{ \Delta{\bf r}\right\}_{\rm diff}} k_{\rm A} \left( \langle A_{{\bf r}+\Delta{\bf r}}\rangle - \langle A_{\bf r}\rangle\right) - \sum_{\left\{ \Delta{\bf r}\right\}_{\rm cat}} \left( k_+\langle A_{\bf r}\rangle -k_- \langle B_{\bf r}\rangle\right) + \sum_{\left\{ \Delta{\bf r}\right\}_{\rm res}} k_{\rm A} \left( \bar{A} - \langle A_{\bf r}\rangle\right)  , \label{eq-av-A}\\
\frac{d}{dt} \langle B_{\bf r}\rangle &=& \sum_{\left\{ \Delta{\bf r}\right\}_{\rm diff}} k_{\rm B} \left( \langle B_{{\bf r}+\Delta{\bf r}}\rangle - \langle B_{\bf r}\rangle\right) + \sum_{\left\{ \Delta{\bf r}\right\}_{\rm cat}} \left( k_+\langle A_{\bf r}\rangle - k_- \langle B_{\bf r}\rangle\right) + \sum_{\left\{ \Delta{\bf r}\right\}_{\rm res}} k_{\rm B} \left( \bar{B} - \langle B_{\bf r}\rangle\right) , \label{eq-av-B}\\
\frac{d}{dt} \langle n\rangle &=& \sum_{\bf r} \sum_{\left\{ \Delta{\bf r}\right\}_{\rm cat}} \left( k_+\langle A_{\bf r}\rangle - k_- \langle B_{\bf r}\rangle\right) .
\label{eq-av-n}
\eea
In Eqs.~(\ref{eq-av-A}) and~(\ref{eq-av-B}), the sums over $\left\{ \Delta{\bf r}\right\}_{\rm cat}$ and $\left\{ \Delta{\bf r}\right\}_{\rm res}$ are possibly vanishing if the cell located at $\bf r$ is not  in contact with the catalytic surface or the reservoir.

If the mean numbers are larger than unity, the fluctuations around the mean values become Gaussian. In this limit, the Markov jump process described by the master equation can be transformed into a diffusive process described by a Fokker-Planck equation by expanding the raising and lowering operators up to second order in the partial derivatives~\cite{G05}. In this way, we can obtain the stochastic partial differential equations (\ref{diff-eq-A})-(\ref{diff-eq-B}) with the boundary conditions (\ref{bc-cat})-(\ref{bc-res}) and the Gaussian white noises~(\ref{Gaussian-eta}) and~(\ref{Gaussian-xi}).

\subsubsection{Equation for the moment generating function}

In order to solve the master equation, we introduce with Gardiner~\cite{G04} the moment generating function,
\be
G\left(z,\{x_{\bf r}\},\{y_{\bf r}\},t\right) \equiv \sum_{n,\{A_{\bf r}\}, \{B_{\bf r}\}}  z^n \, \prod_{\bf r} x_{\bf r}^{A_{\bf r}} \, \prod_{\bf r} y_{\bf r}^{B_{\bf r}} \, P\left(n,\{A_{\bf r}\}, \{B_{\bf r}\},t\right) ,
\label{G-dfn}
\ee
where
\be
z={\rm e}^{-\lambda}
\label{z-lambda}
\ee
and $\lambda$ is the counting parameter.
This generating function obeys the following first-order partial differential equation
\bea
\partial_tG &+& \sum_{\bf r}\biggl\{ \sum_{\left\{ \Delta{\bf r}\right\}_{\rm diff}} \left[ k_{\rm A} \left(x_{\bf r} - x_{{\bf r}+\Delta{\bf r}}\right) \partial_{x_{\bf r}}G  +  k_{\rm B} \left(y_{\bf r} - y_{{\bf r}+\Delta{\bf r}}\right) \partial_{y_{\bf r}}G \right]\nonumber\\
&& \ \quad + \sum_{\left\{ \Delta{\bf r}\right\}_{\rm cat}} \left[ k_+ \left(x_{\bf r} - z \, y_{\bf r}\right) \partial_{x_{\bf r}}G + k_- \left(y_{\bf r} - z^{-1} \, x_{\bf r}\right) \partial_{y_{\bf r}}G \right] \nonumber\\
&& \ \quad + \sum_{\left\{ \Delta{\bf r}\right\}_{\rm res}} \left[ k_{\rm A} \left(x_{\bf r} - 1\right) \partial_{x_{\bf r}}G + k_{\rm B} \left(y_{\bf r} - 1\right) \partial_{y_{\bf r}}G \right]\biggr\}\nonumber\\
&=&  \sum_{\bf r} \ \ \sum_{\left\{ \Delta{\bf r}\right\}_{\rm res}} \left[ k_{\rm A}\bar{A} \left(x_{\bf r} - 1\right) + k_{\rm B}\bar{B} \left(y_{\bf r} - 1\right)\right] \, G \, .
\label{PDE-G}
\eea

Setting
\be
{\bf s} \equiv \left(\{x_{\bf r}\},\{y_{\bf r}\}\right) ,
\label{s-dfn}
\ee
Eq.~(\ref{PDE-G}) can be written as
\be
\partial_t G + \left({\boldsymbol{\mathsf L}} \cdot {\bf s} + {\bf f}\right) \cdot \partial_{\bf s}G = \left({\bf g}\cdot{\bf s}+h\right) G \, , \label{PDE-G-2}
\ee
where
\bea
\left({\boldsymbol{\mathsf L}} \cdot {\bf s}\right) \cdot \pmb{\chi} &\equiv&
 \sum_{\bf r}\biggl\{ \sum_{\left\{ \Delta{\bf r}\right\}_{\rm diff}} \left[ k_{\rm A} \left(x_{\bf r} - x_{{\bf r}+\Delta{\bf r}}\right) \alpha_{\bf r}  +  k_{\rm B} \left(y_{\bf r} - y_{{\bf r}+\Delta{\bf r}}\right) \beta_{\bf r} \right]\nonumber\\
&& \ \quad + \sum_{\left\{ \Delta{\bf r}\right\}_{\rm cat}} \left[ k_+ \left(x_{\bf r} - z \, y_{\bf r}\right) \alpha_{\bf r} + k_- \left(y_{\bf r} - z^{-1} \, x_{\bf r}\right) \beta_{\bf r} \right]  \nonumber\\
&& \ \quad + \sum_{\left\{ \Delta{\bf r}\right\}_{\rm res}} \big( k_{\rm A}\, x_{\bf r} \, \alpha_{\bf r} + k_{\rm B}\, y_{\bf r} \, \beta_{\bf r} \big)\biggr\} ,
\label{sLdG}
\eea
\be
{\bf f} \cdot \pmb{\chi} \equiv - \sum_{\bf r}\sum_{\left\{ \Delta{\bf r}\right\}_{\rm res}} \big( k_{\rm A} \, \alpha_{\bf r} + k_{\rm B} \, \beta_{\bf r} \big) ,
\label{f-dfn}
\ee
\be
{\bf g}\cdot{\bf s} \equiv \sum_{\bf r} \sum_{\left\{ \Delta{\bf r}\right\}_{\rm res}} \big( k_{\rm A}\bar{A} \, x_{\bf r}  + k_{\rm B}\bar{B} \, y_{\bf r} \big) ,
\label{g-dfn}
\ee
and
\be
h \equiv - \sum_{\bf r} \sum_{\left\{ \Delta{\bf r}\right\}_{\rm res}} \big( k_{\rm A}\bar{A}  + k_{\rm B}\bar{B}  \big) ,
\label{h-dfn}
\ee
with the arbitrary vector
\be
\pmb{\chi}  \equiv \left(\{\alpha_{\bf r}\},\{\beta_{\bf r}\}\right) .
\label{chi}
\ee

\subsubsection{Solving the equation for the moment generating function}

As a first-order partial differential equation, Eq.~(\ref{PDE-G-2}) can be solved by the method of characteristics \cite{G04}.  The equations for the characteristics are given by
\bea
&&\frac{d{\bf s}}{dt} = {\boldsymbol{\mathsf L}} \cdot {\bf s} + {\bf f}  \, ,\label{eq-s}\\
&&\frac{dG}{dt} = ({\bf g}\cdot{\bf s}+h)\, G \, ,\label{eq-G}
\eea
where the matrix ${\boldsymbol{\mathsf L}}$ defined by Eq.~(\ref{sLdG}) contains the rate constants and depends on $z$.
Setting $\lambda=0$ and thus $z=1$ in this matrix defines the matrix ${\boldsymbol{\mathsf L}}_0$
such that the kinetic equations~(\ref{eq-av-A}) and~(\ref{eq-av-B}) together read
\be
\frac{d{\pmb{\Gamma}}}{dt} = {\boldsymbol{\mathsf L}}_0^{\rm T} \cdot\big(\pmb{\Gamma}_0-\pmb{\Gamma}\big) \, ,
\label{kin-eqs}
\ee
where
\be
\pmb{\Gamma} = \left(\{\langle A_{\bf r}\rangle\}, \{\langle B_{\bf r}\rangle\}\right)
\ee
are the mean molecular numbers.  The stationary values of these molecular numbers are given by
\be
{\bf\Gamma}_0={\boldsymbol{\mathsf L}}_0^{-1{\rm T}}\cdot{\bf g}
\label{st-st}
\ee
in terms of the vector $\bf g$ defined by Eq.~(\ref{g-dfn}) and containing the elements with the boundary values $\bar{A}$ and $\bar{B}$.  Moreover, the vector $\bf f$ can be written as
\be
{\bf f} = - {\boldsymbol{\mathsf L}}_0 \cdot{\bf 1} \, ,
\label{f-L0}
\ee
which follows by comparing its definition~(\ref{f-dfn}) with Eq.~(\ref{sLdG}) after setting ${\bf s}={\bf 1}$ and $z=1$.  Similarly, the coefficient~(\ref{h-dfn}) is given by
\be
h=-{\bf g}\cdot{\bf 1}\, .
\ee

The solution of Eq.~(\ref{eq-s}) gives the characteristics
\be
{\bf s} = {\rm e}^{{\boldsymbol{\mathsf L}}t}\cdot \left[ {\bf s}_0 + {\boldsymbol{\mathsf L}}^{-1}\cdot\left({\boldsymbol{\mathsf I}}-{\rm e}^{-{\boldsymbol{\mathsf L}}t}\right)\cdot{\bf f}\right] ,
\ee
while the solution of Eq.~(\ref{eq-G}) is given by
\be
G=G_0\, \exp\left[ {\bf g}\cdot{\boldsymbol{\mathsf L}}^{-1}\cdot\left({\boldsymbol{\mathsf I}}-{\rm e}^{-{\boldsymbol{\mathsf L}}t}\right)\cdot\left({\bf s}+{\boldsymbol{\mathsf L}}^{-1}\cdot{\bf f}\right) +\left(h-{\bf g}\cdot{\boldsymbol{\mathsf L}}^{-1}\cdot{\bf f}\right) t\right] .
\ee
The initial condition is the Poisson distribution describing the steady state of Eq.~(\ref{kin-eqs}) and the counter reset to zero $n=0$, so that
\be
G_0(z,{\bf s}_0) = {\rm e}^{{\bf\Gamma}_0\cdot({\bf s}_0-{\bf 1})}
\ee
with the vector~(\ref{st-st}) of the stationary mean values of the molecular numbers.  The solution of the partial differential equation~(\ref{PDE-G}) is thus equal to
\bea
G(z,{\bf s},t) &=& \exp\left[ {\bf g}\cdot{\boldsymbol{\mathsf L}}^{-1}\cdot\left({\boldsymbol{\mathsf I}}-{\rm e}^{-{\boldsymbol{\mathsf L}}t}\right)\cdot\left({\bf s}+{\boldsymbol{\mathsf L}}^{-1}\cdot{\bf f}\right) +\left(h-{\bf g}\cdot{\boldsymbol{\mathsf L}}^{-1}\cdot{\bf f}\right) t\right] \nonumber\\
&\times& \exp\left\{ {\bf\Gamma}_0\cdot\left[ {\rm e}^{-{\boldsymbol{\mathsf L}}t}\cdot{\bf s}-{\boldsymbol{\mathsf L}}^{-1}\cdot\left({\boldsymbol{\mathsf I}}-{\rm e}^{-{\boldsymbol{\mathsf L}}t}\right)\cdot{\bf f}-{\bf 1}\right]\right\} .
\label{G(z,s,t)}
\eea

\subsubsection{Obtaining the cumulant generating function}

We notice that the moment generating function of the number $n$ of reactive events is given by
\be
G(z,{\bf 1},t) = \left\langle {\rm e}^{-\lambda n}\right\rangle_t
\ee
because of Eq.~(\ref{z-lambda}).
The cumulant generating function at time $t$ is thus defined as
\be
Q_t(\lambda) \equiv -\frac{1}{t} \ln G\left(z={\rm e}^{-\lambda},{\bf 1},t\right) ,
\label{Q-G}
\ee
so that we find
\be
Q_t(\lambda) = {\bf g}\cdot\left({\bf 1}+{\boldsymbol{\mathsf L}}^{-1}\cdot{\bf f}\right) -\frac{1}{t}\,{\bf g}\cdot\left({\boldsymbol{\mathsf L}}^{-1}-{\boldsymbol{\mathsf L}}_0^{-1}\right)\cdot\left({\boldsymbol{\mathsf I}}-{\rm e}^{-{\boldsymbol{\mathsf L}}t}\right)\cdot\left({\bf 1}+{\boldsymbol{\mathsf L}}^{-1}\cdot{\bf f}\right) ,
\label{Q-t-1}
\ee
which can be written in the form
\be
Q_t(\lambda) = Q_{\infty}(\lambda) - \frac{1}{t} \, \Xi_t(\lambda) \, ,
\label{Q-inf-Xi}
\ee
where
\be
Q_{\infty}(\lambda) = {\bf g}\cdot\left({\bf 1}+{\boldsymbol{\mathsf L}}^{-1}\cdot{\bf f}\right)
\label{Q-inf-1}
\ee
and
\be
\Xi_t(\lambda) = {\bf g}\cdot\left({\boldsymbol{\mathsf L}}^{-1}-{\boldsymbol{\mathsf L}}_0^{-1}\right)\cdot\left({\boldsymbol{\mathsf I}}-{\rm e}^{-{\boldsymbol{\mathsf L}}t}\right)\cdot\left({\bf 1}+{\boldsymbol{\mathsf L}}^{-1}\cdot{\bf f}\right).
\label{Xi-1}
\ee

We notice that Eq.~(\ref{Xi-1}) converges exponentially towards a constant in the limit $t\to\infty$ if the matrix ${\boldsymbol{\mathsf L}}$ is supposed to be positive, which can be satisfied for some values of $z$ (or $\lambda$) since the rate constants~(\ref{rates}) are positive.

\subsubsection{The dependence of the cumulant generating function on the counting parameter}

A further observation is that
\be
{\boldsymbol{\mathsf L}} = {\boldsymbol{\mathsf M}}\cdot{\boldsymbol{\mathsf L}}_0\cdot{\boldsymbol{\mathsf M}}^{-1}
\label{L-L0-M}
\ee
with
\be
{\boldsymbol{\mathsf M}}\equiv z \, {\boldsymbol{\mathsf P}}_{\rm A} + {\boldsymbol{\mathsf P}}_{\rm B}
\label{M-dfn}
\ee
expressed in terms of the projection matrices
\be
{\boldsymbol{\mathsf P}}_{\rm A} = \left(
\begin{array}{cccccc}
1 & \cdots & 0 & 0 & \cdots & 0 \\
\vdots & \ddots & \vdots & \vdots & \ddots & \vdots \\
0 & \cdots & 1 & 0 & \cdots & 0 \\
0 & \cdots & 0 & 0 & \cdots & 0 \\
\vdots & \ddots & \vdots & \vdots & \ddots & \vdots \\
0 & \cdots & 0 & 0 & \cdots & 0
\end{array}
\right)
\qquad
\mbox{and}
\qquad
{\boldsymbol{\mathsf P}}_{\rm B} = \left(
\begin{array}{cccccc}
0 & \cdots & 0 & 0 & \cdots & 0 \\
\vdots & \ddots & \vdots & \vdots & \ddots & \vdots \\
0 & \cdots & 0 & 0 & \cdots & 0 \\
0 & \cdots & 0 & 1 & \cdots & 0 \\
\vdots & \ddots & \vdots & \vdots & \ddots & \vdots \\
0 & \cdots & 0 & 0 & \cdots & 1
\end{array}
\right) ,
\label{projectors}
\ee
respectively onto the variables of species A and those of species B.  These projection matrices satisfy the condition ${\boldsymbol{\mathsf P}}_{\rm A}+{\boldsymbol{\mathsf P}}_{\rm B}={\boldsymbol{\mathsf I}}$.  We thus have
\bea
{\boldsymbol{\mathsf M}} &=& {\boldsymbol{\mathsf I}} + (z-1) \, {\boldsymbol{\mathsf P}}_{\rm A}={\boldsymbol{\mathsf I}} + \left({\rm e}^{-\lambda}-1\right) \, {\boldsymbol{\mathsf P}}_{\rm A} \, ,\label{M}\\
{\boldsymbol{\mathsf M}}^{-1} &=& {\boldsymbol{\mathsf I}} + (z^{-1}-1) \, {\boldsymbol{\mathsf P}}_{\rm A} ={\boldsymbol{\mathsf I}} + \left({\rm e}^{\lambda}-1\right) \, {\boldsymbol{\mathsf P}}_{\rm A}\, .\label{M-1}
\eea
Therefore, the cumulant generating function can be written as
\be
Q_t(\lambda) = {\bf g}\cdot\left[{\boldsymbol{\mathsf I}}-{\boldsymbol{\mathsf M}}\cdot{\boldsymbol{\mathsf L}}_0^{-1}\cdot{\boldsymbol{\mathsf M}}^{-1}\cdot{\boldsymbol{\mathsf L}}_0 -\frac{1}{t}\left({\boldsymbol{\mathsf M}}\cdot{\boldsymbol{\mathsf L}}_0^{-1}-{\boldsymbol{\mathsf L}}_0^{-1}\cdot{\boldsymbol{\mathsf M}}\right)\cdot\left({\boldsymbol{\mathsf I}}-{\rm e}^{-{\boldsymbol{\mathsf L}}_0t}\right)\cdot\left({\boldsymbol{\mathsf M}}^{-1}-{\boldsymbol{\mathsf L}}_0^{-1}\cdot{\boldsymbol{\mathsf M}}^{-1}\cdot{\boldsymbol{\mathsf L}}_0\right)\right]\cdot{\bf 1}\, .
\ee
As a consequence of Eqs.~(\ref{M}) and~(\ref{M-1}), the previous expression becomes
\bea
Q_t(\lambda) &=& {\bf g}\cdot\biggl[\left(1-z\right){\boldsymbol{\mathsf P}}_{\rm A} + \left(1-z^{-1}\right){\boldsymbol{\mathsf L}}_0^{-1}\cdot{\boldsymbol{\mathsf P}}_{\rm A}\cdot{\boldsymbol{\mathsf L}}_0 - \left(2-z-z^{-1}\right){\boldsymbol{\mathsf P}}_{\rm A} \cdot{\boldsymbol{\mathsf L}}_0^{-1}\cdot{\boldsymbol{\mathsf P}}_{\rm A}\cdot{\boldsymbol{\mathsf L}}_0 \nonumber\\
&& +\frac{1}{t}\left(2-z-z^{-1}\right) \left({\boldsymbol{\mathsf L}}_0^{-1}\cdot{\boldsymbol{\mathsf P}}_{\rm A}-{\boldsymbol{\mathsf P}}_{\rm A}\cdot{\boldsymbol{\mathsf L}}_0^{-1}\right)\cdot\left({\boldsymbol{\mathsf I}}-{\rm e}^{-{\boldsymbol{\mathsf L}}_0t}\right)\cdot\left({\boldsymbol{\mathsf P}}_{\rm A}-{\boldsymbol{\mathsf L}}_0^{-1}\cdot{\boldsymbol{\mathsf P}}_{\rm A}\cdot{\boldsymbol{\mathsf L}}_0\right)\biggr]\cdot{\bf 1}\, .
\eea
Because of Eq.~(\ref{st-st}) and since ${\boldsymbol{\mathsf P}}_{\rm B}={\boldsymbol{\mathsf I}}-{\boldsymbol{\mathsf P}}_{\rm A}$, the cumulant generating function has the form~(\ref{Q}) with the rates
\bea
W_t^{(+)} &=& \pmb{\Gamma}_0 \cdot{\boldsymbol{\mathsf L}}_0\cdot {\boldsymbol{\mathsf P}}_{\rm A} \cdot{\boldsymbol{\mathsf L}}_0^{-1}\cdot{\boldsymbol{\mathsf P}}_{\rm B}\cdot{\boldsymbol{\mathsf L}}_0\cdot{\bf 1} +\frac{1}{t}\, \Psi(t) \, , \\
W_t^{(-)} &=& \pmb{\Gamma}_0 \cdot{\boldsymbol{\mathsf L}}_0\cdot {\boldsymbol{\mathsf P}}_{\rm B}\cdot{\boldsymbol{\mathsf L}}_0^{-1}\cdot{\boldsymbol{\mathsf P}}_{\rm A}\cdot{\boldsymbol{\mathsf L}}_0\cdot{\bf 1} +\frac{1}{t}\, \Psi(t) \, ,
\eea
where
\be
\Psi(t) \equiv \pmb{\Gamma}_0 \cdot\left({\boldsymbol{\mathsf P}}_{\rm A}-{\boldsymbol{\mathsf L}}_0\cdot {\boldsymbol{\mathsf P}}_{\rm A}\cdot{\boldsymbol{\mathsf L}}_0^{-1}\right)\cdot\left({\boldsymbol{\mathsf I}}-{\rm e}^{-{\boldsymbol{\mathsf L}}_0t}\right)\cdot\left({\boldsymbol{\mathsf P}}_{\rm A}-{\boldsymbol{\mathsf L}}_0^{-1}\cdot{\boldsymbol{\mathsf P}}_{\rm A}\cdot{\boldsymbol{\mathsf L}}_0\right)\cdot{\bf 1} \, .
\label{Psi2}
\ee
We have thus proved that the cumulant generating function has the form~(\ref{Q}) and we have obtained explicit expressions for the rates~(\ref{W+})-(\ref{W-}) and the function~(\ref{Psi}) for a discretized space.

\subsubsection{Deducing the probability distribution and its finite-time symmetry}

According to Eq.~(\ref{Q-G}) and the previous results, the moment generating function has the following expression
\be
G(z,{\bf 1},t) = \sum_{n=-\infty}^{+\infty} z^n P(n,t) = {\rm e}^{tW_t^{(+)}(z-1)+tW_t^{(-)}(z^{-1}-1)}
\ee
with the probability distribution
\be
P(n,t) \equiv \sum_{\{A_{\bf r}\}, \{B_{\bf r}\}}  P\left(n,\{A_{\bf r}\}, \{B_{\bf r}\},t\right)
\ee
for the number $n$ of reactive events during the time interval $[0,t]$.
As shown in Ref.~\cite{AG08}, we can use the generating series of Bessel functions given by Eq.~(9.6.33) of Ref.~\cite{AS72},
\be
{\rm e}^{u(q+q^{-1})/2} = \sum_{n=-\infty}^{+\infty} q^n \, I_n(u) \qquad\mbox{for}\qquad q\neq 0 \, .
\ee
Taking
\bea
u &=& 2t\sqrt{W_t^{(+)}W_t^{(-)}} \, , \\
q &=& z\sqrt{\frac{W_t^{(+)}}{W_t^{(-)}}} \, ,
\eea
we get Eq.~(\ref{P(n)}) in Sec.~\ref{Sec-thm}, hence the finite-time fluctuation theorem~(\ref{FT}). Q.E.D.

\subsection{The continuum limit}

\subsubsection{The mean concentrations}

In the continuum limit, we should recover the noiseless diffusion equations (\ref{diff-eq-A})-(\ref{diff-eq-B}) with the noiseless boundary conditions (\ref{bc-cat})-(\ref{bc-res}) for the mean concentrations $\langle c_{\rm A}\rangle$ and $\langle c_{\rm B}\rangle$.  To obtain this result, we introduce the notations
\be
a_{\bf r} \equiv \langle A_{\bf r}\rangle/\Delta r^d \, , \qquad b_{\bf r} \equiv\langle B_{\bf r}\rangle/\Delta r^d \, ,
\ee
where $\Delta V=\Delta r^d$ is the volume element, and we consider Eq.~(\ref{eq-av-A}) for a cell in the bulk of the domain $V$, in which case there is diffusion with all the $2d$ next-neighboring cells and $\{\Delta{\bf r}\}=\{\Delta{\bf r}\}_{\rm diff}$.  Consequently, Eq.~(\ref{eq-av-A}) gives
\be
\frac{da_{\bf r}}{dt}  = \frac{D_{\rm A}}{\Delta r^2}  \sum_{\left\{ \Delta{\bf r}\right\}} \left(a_{{\bf r}+\Delta{\bf r}} - a_{\bf r}\right) ,
\ee
which is the discrete version of the diffusion equation
\be
\partial_t \langle c_{\rm A}\rangle = D_{\rm A} \nabla^2 \langle c_{\rm A}\rangle
\ee
for the mean concentration of species A, $\langle c_{\rm A}({\bf r},t)\rangle=\lim_{\Delta r\to 0} a_{\bf r}(t)=\lim_{\Delta r\to 0} \langle A_{\bf r}(t)\rangle/\Delta r^d$.
Similarly, we get
\be
\partial_t \langle c_{\rm B}\rangle = D_{\rm B} \nabla^2 \langle c_{\rm B}\rangle \, .
\ee

Next, we consider Eq.~(\ref{eq-av-A}) for a cell in contact with the catalyst by the facets $\{\Delta{\bf r}\}_{\rm cat}$.  Therefore, $\{\Delta{\bf r}\}_{\rm diff}=\{\Delta{\bf r}\}\setminus\{\Delta{\bf r}\}_{\rm cat}$ and we find
\be
\frac{da_{\bf r}}{dt}  = \frac{D_{\rm A}}{\Delta r^2} \sum_{\left\{ \Delta{\bf r}\right\}} \left(a_{{\bf r}+\Delta{\bf r}} - a_{\bf r}\right) - \frac{D_{\rm A}}{\Delta r^2} \sum_{\left\{ \Delta{\bf r}\right\}_{\rm cat} } \left(a_{{\bf r}+\Delta{\bf r}} - a_{\bf r}\right)-\frac{1}{\Delta r} \sum_{\left\{ \Delta{\bf r}\right\}_{\rm cat}} \left(\kappa_+ a_{\bf r} - \kappa_- b_{\bf r}\right) .
\ee
As before, the first term gives the discrete version of the Laplacian, while the second can be approximated using
\be
a_{{\bf r}+\Delta{\bf r}} \simeq  a_{\bf r} + \Delta{\bf r}\cdot\pmb{\nabla} a_{\bf r} = a_{\bf r} - \Delta r \, {\bf 1}_{\bot}\cdot\pmb{\nabla} a_{\bf r} ,
\label{grad-a}
\ee
where, as noted earlier, ${\bf 1}_{\bot}$ is the unit vector normal to the surface and oriented towards the interior of the volume~$V$.  For this cell, we thus have
\be
\partial_t  a_{\bf r}  \simeq D_{\rm A}\, \nabla^2a_{\bf r} + \frac{1}{\Delta r}\sum_{\left\{ \Delta{\bf r}\right\}_{\rm cat} } \left[D_{\rm A} \, {\bf 1}_{\bot}\cdot\pmb{\nabla} a-\left(\kappa_+ a - \kappa_- b\right)\right]_{\bf r} ,
\ee
where the derivative $da_{\bf r}/dt$ becomes the partial derivative $\partial_t a_{\bf r}$.
In the limit $ \Delta r\to 0$, consistency is established if every diverging term in the right side is vanishing, which yields the boundary condition
\be
D_{\rm A} \, \partial_{\bot} \langle c_{\rm A}\rangle = \kappa_+ \langle c_{\rm A}\rangle  - \kappa_- \langle c_{\rm B}\rangle  \qquad \mbox{if} \quad {\bf r}\in S_{\rm cat} \, ,
\ee
for the mean concentrations, thus recovering Eq.~(\ref{bc-cat}).  Similarly, Eq.~(\ref{eq-av-B}) gives
\be
-D_{\rm B} \, \partial_{\bot} \langle c_{\rm B}\rangle = \kappa_+ \langle c_{\rm A}\rangle  - \kappa_- \langle c_{\rm B}\rangle  \qquad \mbox{if} \quad {\bf r}\in S_{\rm cat} \, .
\ee
 The boundary conditions on an inert surface are recovered by setting the rate constants equal to zero, $\kappa_{\pm}=0$.

If Eq.~(\ref{eq-av-A}) is considered for a cell in contact with the reservoir by the facets $\{\Delta{\bf r}\}_{\rm res}$, we have that $\{\Delta{\bf r}\}_{\rm diff}=\{\Delta{\bf r}\}\setminus\{\Delta{\bf r}\}_{\rm res}$ and
\be
\frac{da_{\bf r}}{dt}  = \frac{D_{\rm A}}{\Delta r^2} \sum_{\left\{ \Delta{\bf r}\right\}} \left(a_{{\bf r}+\Delta{\bf r}} - a_{\bf r}\right) - \frac{D_{\rm A}}{\Delta r^2} \sum_{\left\{ \Delta{\bf r}\right\}_{\rm res} } \left(a_{{\bf r}+\Delta{\bf r}} - a_{\bf r}\right)+\frac{D_{\rm A}}{\Delta r^2} \sum_{\left\{ \Delta{\bf r}\right\}_{\rm res}} \left(\bar{c}_{\rm A} -  a_{\bf r} \right),
\label{dadt-res}
\ee
because $\bar{c}_{\rm A}=\bar{A}/\Delta r^d$.  Since the first term of Eq.~(\ref{dadt-res}) can also be approximated in terms of the Laplacian and the next terms can be grouped together, we obtain
\be
\partial_t a_{\bf r}  \simeq D_{\rm A}\, \nabla^2a_{\bf r} + \frac{D_{\rm A}}{\Delta r^2}\sum_{\left\{ \Delta{\bf r}\right\}_{\rm res} } \left(\bar{c}_{\rm A}-a_{{\bf r}+\Delta{\bf r}}\right) .
\ee
Again, the consistency is established in the limit $ \Delta r\to 0$ if every diverging term in the right side is vanishing, whereupon we find the boundary conditions $a_{{\bf r}+\Delta{\bf r}}=\bar{c}_{\rm A}$, if $\bf r$ is the center of a cell next to the reservoir and $\Delta{\bf r}\in\{\Delta{\bf r}\}_{\rm res}$.  Consequently, we recover the boundary conditions
\be
 \langle c_{\rm A}\rangle = \bar{c}_{\rm A} \qquad \mbox{and} \qquad  \langle c_{\rm B}\rangle = \bar{c}_{\rm B}
\qquad \mbox{if} \quad {\bf r}\in S_{\rm res} \, ,
\ee
which are given by Eq.~(\ref{bc-res}) for the mean concentrations.

In addition, Eq.~(\ref{eq-av-n}) becomes
\be
\frac{d}{dt} \langle n\rangle = \int_{\rm cat} dS \left( \kappa_+ \langle c_{\rm A}\rangle- \kappa_- \langle c_{\rm B}\rangle\right)
\ee
in the limit $\Delta r\to 0$, because the rates are given by Eq.~(\ref{rates}), the surface element is $\Delta S=\Delta r^{d-1}$, and the sum over the cells having some facets $\{\Delta{\bf r}\}_{\rm cat}$ in common with the catalytic surface converges to a surface integral over the catalyst.

The continuum description is thus recovered for the mean concentrations from the stochastic process introduced by spatial discretization.

\subsubsection{The matrix ${\boldsymbol{\mathsf L}}_0$ in the continuum limit}

In order to interpret more precisely the matrix ${\boldsymbol{\mathsf L}}_0$ in the continuum limit, we take the scalar product of the kinetic equation~(\ref{kin-eqs}) with the vector~(\ref{s-dfn}) and use~Eq.~(\ref{st-st}) to get
\be
{\bf s} \cdot \frac{d{\pmb{\Gamma}}}{dt} =-\pmb{\Gamma}\cdot{\boldsymbol{\mathsf L}}_0\cdot {\bf s}+{\bf g}\cdot{\bf s} \, .
\ee
With the notation $\pmb{\Gamma} = \left(\{a_{\bf r}\Delta r^d\}, \{b_{\bf r}\Delta r^d\}\right)$, Eq.~(\ref{sLdG}) for $z=1$, and Eq.~(\ref{g-dfn}), we obtain
\bea
{\bf s} \cdot \frac{d{\pmb{\Gamma}}}{dt} &=& \sum_{\bf r} \Delta r^d \left( x_{\bf r} \, \frac{da_{\bf r}}{dt} + y_{\bf r} \, \frac{db_{\bf r}}{dt}\right) \nonumber\\
&=&
-\sum_{\bf r}\Delta r^d\biggl\{ \sum_{\left\{ \Delta{\bf r}\right\}_{\rm diff}} \left[ k_{\rm A} \left(x_{\bf r} - x_{{\bf r}+\Delta{\bf r}}\right) a_{\bf r}  +  k_{\rm B} \left(y_{\bf r} - y_{{\bf r}+\Delta{\bf r}}\right) b_{\bf r} \right]\nonumber\\
&& \qquad \qquad + \sum_{\left\{ \Delta{\bf r}\right\}_{\rm cat}} \left(x_{\bf r} - y_{\bf r}\right) \left( k_+  a_{\bf r} - k_-  b_{\bf r} \right)  \nonumber\\
&& \qquad \qquad + \sum_{\left\{ \Delta{\bf r}\right\}_{\rm res}} \left[ k_{\rm A}x_{\bf r} \, \left(a_{\bf r}-\bar{c}_{\rm A}\right) + k_{\rm B}\, y_{\bf r} \left( b_{\bf r}-\bar{c}_{\rm B}\right) \right]\biggr\} .
\label{sdgdt-1}
\eea
Using the identity
\be
\sum_{\bf r}\sum_{\left\{ \Delta{\bf r}\right\}_{\rm diff}} x_{{\bf r}+\Delta{\bf r}} \, a_{\bf r} = \sum_{\bf r}\sum_{\left\{ \Delta{\bf r}\right\}_{\rm diff}} x_{\bf r} \, a_{{\bf r}+\Delta{\bf r}} \, ,
\ee
a similar relation for $y_{{\bf r}+\Delta{\bf r}}$ and~$b_{\bf r}$, as well as $\{\Delta{\bf r}\}_{\rm diff}=\{\Delta{\bf r}\}\setminus\{\Delta{\bf r}\}_{\rm cat}\setminus\{\Delta{\bf r}\}_{\rm res}$, Eq.~(\ref{sdgdt-1}) becomes
\bea
{\bf s} \cdot \frac{d{\pmb{\Gamma}}}{dt} &=& \sum_{\bf r} \Delta r^d\left( x_{\bf r} \, \frac{da_{\bf r}}{dt} + y_{\bf r} \, \frac{db_{\bf r}}{dt}\right) \nonumber\\
&=&
\sum_{\bf r}\Delta r^d\sum_{\left\{ \Delta{\bf r}\right\}} \left[ k_{\rm A} \, x_{\bf r} \left(a_{{\bf r}+\Delta{\bf r}}-a_{\bf r}\right)  +  k_{\rm B} \, y_{\bf r} \left(b_{{\bf r}+\Delta{\bf r}}-b_{\bf r}\right) \right]\nonumber\\
&-& \sum_{\bf r}\Delta r^d\sum_{\left\{ \Delta{\bf r}\right\}_{\rm cat}} \left[ k_{\rm A} \, x_{\bf r} \left(a_{{\bf r}+\Delta{\bf r}}-a_{\bf r}\right)  +  k_{\rm B} \, y_{\bf r} \left(b_{{\bf r}+\Delta{\bf r}}-b_{\bf r}\right) +\left(x_{\bf r} - y_{\bf r}\right) \left( k_+  a_{\bf r} - k_-  b_{\bf r} \right)\right]  \nonumber\\
&-& \sum_{\bf r}\Delta r^d\sum_{\left\{ \Delta{\bf r}\right\}_{\rm res}} \left[ k_{\rm A}x_{\bf r} \, \left(a_{{\bf r}+\Delta{\bf r}}-\bar{c}_{\rm A}\right) + k_{\rm B}\, y_{\bf r} \left( b_{{\bf r}+\Delta{\bf r}}-\bar{c}_{\rm B}\right) \right] .
\label{sdgdt-2}
\eea
Substituting the expressions~(\ref{rates}) for the rates, and using the approximations~(\ref{grad-a}), we have that
\bea
\sum_{\bf r} \Delta r^d\left( x_{\bf r} \, \frac{da_{\bf r}}{dt} + y_{\bf r} \, \frac{db_{\bf r}}{dt}\right)
&\simeq&
\sum_{\bf r}\Delta r^d \left(D_{\rm A} \, x_{\bf r} \, \nabla^2 a_{\bf r}  +  D_{\rm B} \, y_{\bf r} \, \nabla^2 b_{\bf r}\right)\nonumber\\
&+& \sum_{\bf r}\Delta r^{d-1} \sum_{\left\{ \Delta{\bf r}\right\}_{\rm cat}} \left[ D_{\rm A} \, x_{\bf r} \, {\bf 1}_{\bot}\cdot\pmb{\nabla} a_{\bf r}  +  D_{\rm B} \, y_{\bf r} \, {\bf 1}_{\bot}\cdot\pmb{\nabla} b_{\bf r} -\left(x_{\bf r} - y_{\bf r}\right) \left( \kappa_+  a_{\bf r} - \kappa_-  b_{\bf r} \right)\right]  \nonumber\\
&-& \sum_{\bf r}\Delta r^{d-2} \sum_{\left\{ \Delta{\bf r}\right\}_{\rm res}} \left[ D_{\rm A}x_{\bf r} \, \left(a_{{\bf r}+\Delta{\bf r}}-\bar{c}_{\rm A}\right) + D_{\rm B}\, y_{\bf r} \left( b_{{\bf r}+\Delta{\bf r}}-\bar{c}_{\rm B}\right) \right] .
\label{sdgdt-3}
\eea
In the limit $\Delta r\to 0$, the last terms at the boundary with the reservoir are vanishing because of the boundary conditions $a_{{\bf r}+\Delta{\bf r}}=\bar{c}_{\rm A}$ and $b_{{\bf r}+\Delta{\bf r}}=\bar{c}_{\rm B}$ and we find
\bea
\int dV \left( x \, \partial_t a + y \, \partial_t b\right)
&=&
\int dV \left( x \, D_{\rm A} \, \nabla^2 a  +  y \, D_{\rm B} \, \nabla^2 b\right)\nonumber\\
&+& \int_{\rm cat} dS \left[ x \left( D_{\rm A} \, \partial_{\bot} a - \kappa_+  a + \kappa_-  b \right) +  y \left( D_{\rm B} \, \partial_{\bot}b + \kappa_+  a - \kappa_-  b \right)\right]   .
\label{sdgdt-4}
\eea
We notice that the diffusion equations and the reactive boundary conditions are recovered by considering variations of this equation with respect to $x$ and $y$.  Therefore, the matrix ${\boldsymbol{\mathsf L}}_0$ can be interpreted as the evolution operator  of the diffusion equations combined with the boundary conditions of the problem.  The result is consistent with the fact that Eq.~(\ref{kin-eqs}) corresponds to the macroscopic diffusion equations.

\subsubsection{The asymptotic cumulant generating function}

Here, we calculate the asymptotic value~(\ref{Q-inf-1}) of the cumulant generating function~(\ref{Q-t-1}).
Denoting $\pmb{\gamma}$ the solution of the problem
\be
{\boldsymbol{\mathsf L}}^{{\rm T}}\cdot\pmb{\gamma}={\bf g} \, ,
\label{chi-L-g}
\ee
and using Eq.~(\ref{f-L0}), the asymptotic cumulant generating function can be expressed as
\be
Q_{\infty}(\lambda) = \pmb{\gamma}\cdot\left({\boldsymbol{\mathsf L}}-{\boldsymbol{\mathsf L}}_0\right)\cdot{\bf 1} \, .
\label{Q-L-L0}
\ee
With the same method as before and replacing in Eq.~(\ref{sLdG}) the vector $\pmb{\chi}$ by
\be
\pmb{\gamma}=(\{\tilde a_{\bf r}\Delta r^d\},\{\tilde b_{\bf r}\Delta r^d\}) \, ,
\label{gamma-tilde}
\ee
we get
\bea
\pmb{\gamma}\cdot{\boldsymbol{\mathsf L}} \cdot {\bf s} &\simeq&
-\int dV \left( x \, D_{\rm A} \, \nabla^2 \tilde a  +  y \, D_{\rm B} \, \nabla^2 \tilde b\right)\nonumber\\
&& -\int_{\rm cat} dS \left[ x \left( D_{\rm A} \, \partial_{\bot} \tilde a - \kappa_+  \tilde a + z^{-1} \kappa_-  \tilde b \right) +  y \left( D_{\rm B} \, \partial_{\bot}\tilde b + z \kappa_+  \tilde a - \kappa_-  \tilde b \right)\right] \nonumber\\
&& +\frac{1}{\Delta r}\int_{\rm res} dS \left( x \, D_{\rm A} \, \tilde a +  y \, D_{\rm B} \, \tilde b \right) ,
\label{sLdG-c}
\eea
while Eq.~(\ref{g-dfn}) becomes
\be
{\bf g}\cdot{\bf s} \simeq \frac{1}{\Delta r}\int_{\rm res} dS \left( x \, D_{\rm A} \, \bar{c}_{\rm A} +  y \, D_{\rm B} \, \bar{c}_{\rm B} \right) .
\label{g-s-c}
\ee
Since Eq.~(\ref{chi-L-g}) implies the equality $\pmb{\gamma}\cdot{\boldsymbol{\mathsf L}} \cdot {\bf s} ={\bf g}\cdot{\bf s}$ for any vector $\bf s$, its solution can be expressed in terms of the fields $\tilde a({\bf r})\equiv \lim_{\Delta r\to 0} \tilde a_{\bf r}$ and $\tilde b({\bf r})\equiv \lim_{\Delta r\to 0} \tilde b_{\bf r}$ that are given by solving
\bea
&&\nabla^2  \tilde{a} = 0 \, , \label{ctilde1}\\
&&\nabla^2  \tilde{b} = 0 \, , \label{ctilde2}\\
&&D_{\rm A}\left(\partial_{\bot} \tilde a\right)_{\rm cat} = \left( \kappa_+ \tilde a -z^{-1} \kappa_- \tilde b \right)_{\rm cat} \, , \label{ctilde3}\\
&&D_{\rm B}\left(\partial_{\bot} \tilde b\right)_{\rm cat} = -\left( z\,\kappa_+ \tilde a - \kappa_- \tilde b \right)_{\rm cat} \, , \label{ctilde4}\\
&& \left(\tilde a\right)_{\rm res} = \bar{c}_{\rm A} \, , \label{ctilde5}\\
&& \left(\tilde b\right)_{\rm res} = \bar{c}_{\rm B} \, . \label{ctilde6}
\eea
Setting
\bea
\tilde a({\bf r}) &=& \bar{c}_{\rm A} - \frac{\ell}{D_{\rm A}} \left( \kappa_+ \bar{c}_{\rm A} -z^{-1} \kappa_- \bar{c}_{\rm B}\right) \phi({\bf r}) \, , \label{tilde-c-A}\\
\tilde b({\bf r}) &=& \bar{c}_{\rm B} + \frac{\ell}{D_{\rm B}} \left( z\, \kappa_+ \bar{c}_{\rm A} - \kappa_- \bar{c}_{\rm B}\right) \phi({\bf r}) \, ,  \label{tilde-c-B}
\eea
we find that the field $\phi({\bf r})$ is the solution of Eqs.~(\ref{phi1})-(\ref{phi4}).

In order to calculate~(\ref{Q-L-L0}), we set ${\bf s}={\bf 1}$ in Eq.~(\ref{sLdG-c}) and substract the same expression with $z=1$.  Since $z={\rm e}^{-\lambda}$, we obtain
\be
Q_{\infty}(\lambda) = \int_{\rm cat} dS\left[ \kappa_+\tilde a \left(1-{\rm e}^{-\lambda}\right) +  \kappa_-\tilde b \left(1-{\rm e}^{\lambda}\right) \right]
\label{Q-inf-ctilde}
\ee
in terms of the solution of Eqs.~(\ref{ctilde1})-(\ref{ctilde6}).  Substituting Eqs.~(\ref{tilde-c-A})-(\ref{tilde-c-B}) therein yields
\be
Q_{\infty}(\lambda) = \int_{\rm cat} dS\, (1-\phi) \left[ \kappa_+\bar{c}_{\rm A} \left(1-{\rm e}^{-\lambda}\right) +  \kappa_-\bar{c}_{\rm B} \left(1-{\rm e}^{\lambda}\right) \right] .
\ee
According to Eq.~(\ref{Sigma}), we thus find
\be
Q_{\infty}(\lambda) = W_{\infty}^{(+)} \left(1-{\rm e}^{-\lambda}\right) + W_{\infty}^{(-)} \left(1-{\rm e}^{\lambda}\right) ,
\label{Q-inf}
\ee
proving that the asymptotic values of the rates~(\ref{W+})-(\ref{W-}) are indeed given by Eqs.~(\ref{Winf+}) and~(\ref{Winf-}).

\subsubsection{The time-dependent contribution to the cumulant generating function}

Here, we calculate the time-dependent function~(\ref{Xi-1}), which appears in the last term of the cumulant generating function~(\ref{Q-t-1}).  Using Eq.~(\ref{L-L0-M}), the function~(\ref{Xi-1}) becomes
\be
\Xi_t(\lambda) = {\bf g}\cdot\left({\boldsymbol{\mathsf L}}^{-1}-{\boldsymbol{\mathsf L}}_0^{-1}\right)\cdot{\boldsymbol{\mathsf M}}\cdot\left({\boldsymbol{\mathsf I}}-{\rm e}^{-{\boldsymbol{\mathsf L}}_0t}\right)\cdot{\boldsymbol{\mathsf M}}^{-1}\cdot\left({\bf 1}+{\boldsymbol{\mathsf L}}^{-1}\cdot{\bf f}\right) .
\label{Xi-2}
\ee
On the one hand, the vector $\bf g$ can be expressed in terms of the stationary state $\pmb{\Gamma}_0$ according to Eq.~(\ref{st-st}), as well as in terms of the vector $\pmb{\gamma}$ given above by Eq.~(\ref{chi-L-g}),
\be
{\bf g}={\boldsymbol{\mathsf L}}_0^{{\rm T}}\cdot\pmb{\Gamma}_0={\boldsymbol{\mathsf L}}^{{\rm T}}\cdot\pmb{\gamma} \, .
\ee
On the other hand, the vector $\bf f$ can be written as in Eq.~(\ref{f-L0}).  Consequently, the function~(\ref{Xi-2}) is of the form
\be
\Xi_t(\lambda) = \pmb{\eta}\cdot\left({\boldsymbol{\mathsf I}}-{\rm e}^{-{\boldsymbol{\mathsf L}}_0t}\right)\cdot\pmb{\xi}
\label{Xi-3}
\ee
with
\be
\pmb{\eta} \equiv \left(\pmb{\gamma}-\pmb{\Gamma}_0\right)\cdot{\boldsymbol{\mathsf M}}
\label{eta-dfn}
\ee
and
\be
\pmb{\xi}\equiv {\boldsymbol{\mathsf L}}_0^{-1}\cdot{\boldsymbol{\mathsf M}}^{-1}\cdot\left({\boldsymbol{\mathsf L}}-{\boldsymbol{\mathsf L}}_0\right)\cdot{\bf 1} \, .
\label{xi-dfn}
\ee
Using Eqs.~(\ref{M}) and~(\ref{M-1}), we notice that
\bea
&&\pmb{\eta}=(1-z)\, \pmb{\Gamma}_0 \cdot\left({\boldsymbol{\mathsf P}}_{\rm A}-{\boldsymbol{\mathsf L}}_0\cdot {\boldsymbol{\mathsf P}}_{\rm A}\cdot{\boldsymbol{\mathsf L}}_0^{-1}\right) , \label{eta2}\\
&&\pmb{\xi} = (z^{-1}-1) \left({\boldsymbol{\mathsf P}}_{\rm A}-{\boldsymbol{\mathsf L}}_0^{-1}\cdot{\boldsymbol{\mathsf P}}_{\rm A}\cdot{\boldsymbol{\mathsf L}}_0\right)\cdot{\bf 1} \, , \label{xi2}
\eea
showing that we should expect the factorizations of $1-z$ and $z^{-1}-1$, respectively.  In order to establish this factorization and obtain the analytical expressions for (\ref{eta-dfn}) and~(\ref{xi-dfn}) in the continuum limit, we proceed as follows.

Multiplying Eq.~(\ref{eta-dfn}) by an arbitrary vector ${\bf s}=\left( \{x_{\bf r}\}, \{y_{\bf r}\}\right)$ and using the notations (\ref{gamma-tilde}),
\be
\pmb{\eta}=\left( \{u_{\bf r}\Delta r^d\}, \{v_{\bf r}\Delta r^d\}\right) \, , \qquad \pmb{\Gamma}_0 = \left(\{\langle A_{\bf r}\rangle_{\rm st}\}, \{\langle B_{\bf r}\rangle_{\rm st}\}\right) ,
\ee
as well as the definition~(\ref{M-dfn}), we have that
\be
\pmb{\eta}\cdot{\bf s}=\sum_{\bf r}\Delta r^d \left(u_{\bf r}\, x_{\bf r} + v_{\bf r}\,  y_{\bf r} \right) = \sum_{\bf r}  \left[ z\left( \tilde a_{\bf r}\Delta r^d-\langle A_{\bf r}\rangle_{\rm st}\right) x_{\bf r} + \left( \tilde b_{\bf r}\Delta r^d -\langle B_{\bf r}\rangle_{\rm st}\right) y_{\bf r} \right] .
\ee
Since
\be
\langle A_{\bf r}\rangle_{\rm st}  = \langle c_{\rm A}({\bf r})\rangle_{\rm st} \, \Delta r^d \qquad\mbox{and}\qquad \langle B_{\bf r}\rangle_{\rm st}  = \langle c_{\rm B}({\bf r})\rangle_{\rm st} \, \Delta r^d \, ,
\ee
Eqs.~(\ref{NESS}), (\ref{tilde-c-A}), and~(\ref{tilde-c-B}) yield
\bea
&& u_{\bf r} = \left(1-z\right) \ell \kappa_- \frac{\bar{c}_{\rm B}}{D_{\rm A}} \, \phi({\bf r})  \, , \label{breve-a}\\
&& v_{\bf r} = -\left(1-z\right) \ell \kappa_+ \frac{\bar{c}_{\rm A}}{D_{\rm B}} \, \phi({\bf r})  \, , \label{breve-b}
\eea
confirming the factorization expected by Eq.~(\ref{eta2}) and expressing~(\ref{eta-dfn}) in terms of the solution $\phi({\bf r})$ of the stationary problem (\ref{phi1})-(\ref{phi4}).

Multiplying Eq.~(\ref{xi-dfn}) by ${\boldsymbol{\mathsf L}}_0$ and an arbitrary vector~(\ref{chi}), we obtain the equation
\be
\pmb{\chi}\cdot{\boldsymbol{\mathsf L}}_0\cdot\pmb{\xi}= \pmb{\chi}\cdot{\boldsymbol{\mathsf M}}^{-1}\cdot\left({\boldsymbol{\mathsf L}}-{\boldsymbol{\mathsf L}}_0\right)\cdot{\bf 1}
\label{xi-eq}
\ee
that the vector $\pmb{\xi}=\left( \{\tilde x_{\bf r}\}, \{\tilde y_{\bf r}\}\right)$ should satisfy.
With the same method as before, we get
\bea
\pmb{\chi}\cdot{\boldsymbol{\mathsf L}}_0 \cdot \pmb{\xi} &\simeq&
-\frac{1}{\Delta r^d} \int dV \left(\tilde x \, D_{\rm A} \, \nabla^2 \alpha  + \tilde y \, D_{\rm B} \, \nabla^2 \beta\right)\nonumber\\
&& -\frac{1}{\Delta r^d}\int_{\rm cat} dS \left[\tilde x \left( D_{\rm A} \, \partial_{\bot} \alpha - \kappa_+  \alpha +  \kappa_-  \beta \right) +\tilde  y \left( D_{\rm B} \, \partial_{\bot}\beta +  \kappa_+  \alpha - \kappa_-  \beta \right)\right] \nonumber\\
&& +\frac{1}{\Delta r^{d+1}}\int_{\rm res} dS \left(\tilde x \, D_{\rm A} \, \alpha +\tilde  y \, D_{\rm B} \, \beta \right) ,
\label{xi-eq-LHS}
\eea
and
\be
\pmb{\chi}\cdot{\boldsymbol{\mathsf M}}^{-1}\cdot\left({\boldsymbol{\mathsf L}}-{\boldsymbol{\mathsf L}}_0\right)\cdot{\bf 1}
\simeq \frac{1}{\Delta r^d}\int_{\rm cat} dS \left(z^{-1}-1\right)\left( \kappa_+  \alpha - \kappa_-  \beta \right) .
\label{xi-eq-RHS}
\ee
At the leading order $1/\Delta r^{d+1}$, the equality~(\ref{xi-eq}) between~(\ref{xi-eq-LHS}) and~(\ref{xi-eq-RHS}) shows that the boundary conditions $(\tilde x)_{\rm res}=(\tilde y)_{\rm res}=0$ should be satisfied on the reservoir.  Now, integrating by parts leads to
\be
\int dV \tilde x \, \nabla^2 \alpha = \int dV \alpha \, \nabla^2 \tilde x + \int dS \left( \alpha\, \partial_{\bot} \tilde x - \tilde x \, \partial_{\bot} \alpha\right)
\ee
and a similar relation between $\tilde y$ and $\beta$.  Accordingly, at the subleading order $1/\Delta r^{d}$, Eq.~(\ref{xi-eq}) becomes
\bea
&&\int dV \left( \alpha \, D_{\rm A} \, \nabla^2 \tilde x  +  \beta \, D_{\rm B} \, \nabla^2\tilde y\right)\nonumber\\
&& +\int_{\rm cat} dS \left\{ \alpha \left[ D_{\rm A} \, \partial_{\bot}\tilde x - \kappa_+  \left(\tilde x-\tilde y+1-z^{-1}\right)\right]
+ \beta \left[ D_{\rm B} \, \partial_{\bot}\tilde y + \kappa_-  \left(\tilde x-\tilde y+1-z^{-1}\right) \right] \right\}
=0 \, .
\eea
Taking variations with respect to $\alpha$ and $\beta$, we find that the fields $\tilde x$ and $\tilde y$ are the solutions of the following problem:
\bea
&& \nabla^2 \tilde x = 0 \, , \\
&& \nabla^2 \tilde y = 0 \, , \\
&& D_{\rm A} \left(\partial_{\bot}\tilde x\right)_{\rm cat} = \kappa_+  \left(\tilde x-\tilde y+1-z^{-1}\right)_{\rm cat} \, , \\
&& D_{\rm B} \left(\partial_{\bot} \tilde y\right)_{\rm cat} = -\kappa_-  \left(\tilde x-\tilde y+1-z^{-1}\right)_{\rm cat} \, , \\
&& (\tilde x)_{\rm res}=0 \, ,\\
&& (\tilde y)_{\rm res}=0 \, .
\eea
With the substitution
\bea
\tilde x({\bf r}) &=& \left(z^{-1}-1\right) \ell \, \frac{\kappa_+}{D_{\rm A}}\, \phi({\bf r}) \, , \label{breve-x}\\
\tilde y({\bf r}) &=& -\left(z^{-1}-1\right) \ell \, \frac{\kappa_-}{D_{\rm B}} \, \phi({\bf r}) \, , \label{breve-y}
\eea
the problem is reduced to finding the solution $\phi({\bf r})$ of Eqs.~(\ref{phi1})-(\ref{phi4}).

Now, Eq.~(\ref{Xi-3}) can be rewritten as
\be
\Xi_t(\lambda) = \pmb{\eta}\cdot\left(\pmb{\xi}-\pmb{\xi}_t\right)
\label{Xi-4}
\ee
in terms of the time-dependent vector
\be
\pmb{\xi}_t = {\rm e}^{-{\boldsymbol{\mathsf L}}_0t}\cdot\pmb{\xi} \, ,
\label{xi_t-dfn}
\ee
which is the solution of
\be
\frac{d\pmb{\xi}_t}{dt} = -{\boldsymbol{\mathsf L}}_0\cdot\pmb{\xi}_t
\label{xi_t-eq}
\ee
and denoted $\pmb{\xi}_t=\left( \{\tilde x_{t,{\bf r}}\}, \{\tilde y_{t,{\bf r}}\}\right)$.
Multiplying Eq.~(\ref{xi_t-eq}) by an arbitrary vector~(\ref{chi}) and using the same method as above, we find that
\bea
\int dV \left(\alpha\, \partial_t \tilde x_t +\beta\, \partial_t \tilde y_t\right) &=& \int dV \left( \alpha \, D_{\rm A} \, \nabla^2 \tilde x_t  +  \beta \, D_{\rm B} \, \nabla^2 \tilde y_t\right)\nonumber\\
&& +\int_{\rm cat} dS \left\{ \alpha \left[ D_{\rm A} \, \partial_{\bot} \tilde x_t - \kappa_+  \left(\tilde x_t-\tilde y_t\right)\right]
+ \beta \left[ D_{\rm B} \, \partial_{\bot} \tilde y + \kappa_-  \left(\tilde x_t-\tilde y_t\right) \right] \right\}
\eea
with the boundary conditions $(\tilde x_t)_{\rm res}=(\tilde y_t)_{\rm res}=0$.  The fields $\tilde x_t$ and $\tilde y_t$ are thus the solutions of the following problem:
\bea
&& \partial_t \tilde x_t = D_{\rm A} \nabla^2 \tilde x_t  \, , \\
&& \partial_t \tilde y_t = D_{\rm B} \nabla^2 \tilde y_t \, , \\
&& D_{\rm A} \left(\partial_{\bot} \tilde x_t\right)_{\rm cat} = \kappa_+  \left(\tilde x_t-\tilde y_t\right)_{\rm cat} \, , \\
&& D_{\rm B} \left(\partial_{\bot} \tilde y_t\right)_{\rm cat} = -\kappa_-  \left(\tilde x_t-\tilde y_t\right)_{\rm cat} \, , \\
&& (\tilde x_t)_{\rm res}=0 \, ,\\
&& (\tilde y_t)_{\rm res}=0 \, , \\
&& \tilde x_{t=0} = \tilde x \, , \\
&& \tilde y_{t=0} = \tilde y \, ,
\eea
where the initial conditions are given in terms of the previously obtained stationary functions $\tilde x$ and $\tilde y$.  Setting
\bea
\tilde x_{t}({\bf r}) &=& \left(z^{-1}-1\right) \ell \, \frac{\kappa_+}{D_{\rm A}} \, f_{\rm A}({\bf r},t) \, , \label{breve-x-t}\\
\tilde y_{t}({\bf r}) &=& -\left(z^{-1}-1\right) \ell \, \frac{\kappa_-}{D_{\rm B}} \, f_{\rm B}({\bf r},t) \, , \label{breve-y-t}
\eea
we conclude that the functions $f_k({\bf r},t)$ obey the time-dependent problem of Eqs.~(\ref{f1})-(\ref{f5}) for $k={\rm A},{\rm B}$. We note that the factorization expected by Eq.~(\ref{xi2}) is confirmed by Eqs.~(\ref{breve-x})-(\ref{breve-y}) and~(\ref{breve-x-t})-(\ref{breve-y-t}).

Substituting the results (\ref{breve-a})-(\ref{breve-b}), (\ref{breve-x})-(\ref{breve-y}), and (\ref{breve-x-t})-(\ref{breve-y-t}) into Eq.~(\ref{Xi-4}), we find
\bea
\Xi_t(\lambda) &=&\sum_{\bf r} \Delta r^d \left[ u_{\bf r}\left(\tilde x_{\bf r}-\tilde x_{t,{\bf r}}\right) + v_{\bf r}\left(\tilde y_{\bf r}-\tilde y_{t,{\bf r}}\right)\right]\nonumber\\
&\simeq& \left(1-z\right)\left(z^{-1}-1\right) \ell^2 \kappa_+\kappa_-  \sum_{\bf r} \Delta r^d \, \phi({\bf r}) \left\{ \frac{\bar{c}_{\rm B}}{D_{\rm A}^2} \left[\phi({\bf r})-f_{\rm A}({\bf r},t)\right] + \frac{\bar{c}_{\rm A}}{D_{\rm B}^2} \left[\phi({\bf r})-f_{\rm B}({\bf r},t)\right]\right\} .
\label{Xi-5}
\eea
Since $z={\rm e}^{-\lambda}$, we finally obtain
\be
\Xi_t(\lambda) =  \left({\rm e}^{\lambda}+{\rm e}^{-\lambda}-2\right) \Psi(t)
\label{Xi}
\ee
expressed in terms of the function~(\ref{Psi}) in the continuum limit $\Delta r\to 0$.

The analytic expressions~(\ref{W+}) and~(\ref{W-}) for the rates are thus proved.

\subsubsection{The cumulant generating function at early time}

Expanding the function~(\ref{Xi-3}) in powers of time, keeping the term of first order in the time $t$, and replacing $Q_\infty(\lambda)$ with its expression~(\ref{Q-L-L0}) in Eq.~(\ref{Q-inf-Xi}), we get the following expression
\be
Q_t(\lambda) = \pmb{\Gamma}_0\cdot\left({\boldsymbol{\mathsf L}}-{\boldsymbol{\mathsf L}}_0\right)\cdot{\bf 1} + O(t)\, ,
\label{Q-Gamma0}
\ee
showing that the early-time behavior of the cumulant generating function is given by an expression similar to Eq.~(\ref{Q-L-L0}), but with the vector $\pmb{\gamma}$ corresponding to the fields $(\tilde a,\tilde b)$ substituted by the stationary state~(\ref{st-st}) corresponding to the fields $(\langle c_{\rm A}\rangle_{\rm st},\langle c_{\rm B}\rangle_{\rm st})$.  Carrying out this substitution in Eq.~(\ref{Q-inf-ctilde}), which is the continuum limit of Eq.~(\ref{Q-L-L0}), we obtain Eq.~(\ref{Q-early}).

\section{Conclusion and perspectives}
\label{conclusion}

In this paper, a finite-time fluctuation theorem was established for the diffusion-influenced surface reaction ${\rm A}\rightleftharpoons{\rm B}$ ruled by stochastic partial differential equations.  The theorem was deduced by solving the evolution equation for the moment generating function of a corresponding spatially discretized system, thereafter taking the continuum limit.  The analytical expression of the cumulant generating function is thus given in terms of the finite-time rates of the diffusion-influenced reaction process.  In this way, the large-deviation properties of the spatially extended stochastic process are obtained by solving deterministic diffusion equations with specific boundary conditions.

The results show that, in stationary states, the full counting statistics of the reactive events satisfies a time-reversal symmetry over every finite time interval, which finds its origin in microreversibility.  The affinity of the fluctuation theorem also depends on time with a known analytical dependence.  In this diffusion-influenced system, one of the prominent features of this affinity is that it may take different values at finite time than its asymptotic value predicted by the standard infinite-time fluctuation theorem.

The finite-time fluctuation theorem holds because the macroscopic rate of the reaction ${\rm A}\rightleftharpoons{\rm B}$ is linear in the concentrations (although nonlinear in the affinity).  Therefore, the generating function~(\ref{G(z,s,t)}) of the joint conditional probability distribution for the numbers of molecules and reactive events remains exponential in the generating variables, $\bf s$, associated with the numbers of molecules, if the counting starts from the Poissonian stationary state.  In this regard, we may conjecture that the result can be extended to networks of diffusion-influenced surface reactions having macroscopic rates that are linearly dependent on the concentrations.  In such systems, several currents may be coupled together, leading to Onsager reciprocal relations and their generalizations to the nonlinear response regimes~\cite{AG04}.

\section*{Acknowledgments}

The Authors thank Patrick Grosfils and Mu-Jie Huang for fruitful discussions. Financial support from the International Solvay Institutes for Physics and Chemistry, the Universit\'e libre de Bruxelles (ULB), the Fonds de la Recherche Scientifique~-~FNRS under the Grant PDR~T.0094.16 for the project ``SYMSTATPHYS", and the Natural Sciences and Engineering Research Council of Canada is acknowledged.





\end{document}